\documentstyle[12pt,leqno]{article}
\newcommand{\qed}{\hfill $\Box$ \\}

\font\msbm=msbm10 at 12pt
\newcommand{\ZZ}{\mbox{\msbm Z}}

\newcommand{\FF}{\mbox{\msbm F}}

\def \Z {{\ZZ}}
\def \z {{\ZZ}}
\def \C {{\cal C}}
\def \x {{\bf x}}
\def \w {{\bf w}}
\def \u {{\bf u}}
\def \v {{\bf v}}
\def \y {{\bf y}}
\def \c {{\bf c}}

\def \M {{\cal M}}

\def \v {{\bf v}}
\def \c {{\bf c}}

\def \v {{\bf v}}
\def \c {{\bf c}}
\def \D {{\cal D}}
\newtheorem{theorem}{Theorem}

\newtheorem{proposition}{Proposition}

\newcommand{\be} {\begin{equation}}
\newcommand{\ee} {\end{equation}}

\begin{document}

\title{On the Covering Radius of Some Modular Codes
\thanks{The work of the second author was done while visiting at DA-IICT Gandhinagar, India.  The second author was 
supported by a grant (SR/S4/MS:588/09) from Department of Science and Technology (DST), India.}
}
\author{Manish.~K.~Gupta\\
Laboratory of Natural Information Processing,\\
Dhirubhai Ambani Institute of Information and Communication Technology, \\
Gandhinagar, Gujarat, 382007 India\\
{Email: m.k.gupta@ieee.org} 
\bigskip\\
C.~Durairajan\\
Department of Mathematics,\\ 
School of Mathematical Sciences\\
Bharathidasan University\\
Tiruchirappalli Tamil Nadu 620 024 India\\
{Email: durairajanc@gmail.com} \\
\hfill \\
\hfill \\
\hfill \\
\hfill \\
{\bf Proposed running head:} On the Covering Radius of Some Modular Codes
}
\date{}
\maketitle

\newpage

\vspace*{0.5cm}
\begin{abstract}
This paper gives lower and upper bounds on the  covering radius of codes 
over $\Z_{2^s}$ with respect to homogenous distance. We also determine the covering radius of various 
Repetition codes, Simplex codes (Type $\alpha$ and Type $\beta$)  
and their dual and give bounds on the covering radii for MacDonald codes of both types over $\Z_4$. 
\end{abstract}
\vspace*{0.5cm}

\vspace{0.5cm}
\vspace{1.5cm}

\noindent
Corresponding author:\\ 
\\
 \hspace{1cm} 
 Manish K. Gupta\\
 \\ 
\noindent
Laboratory of Natural Information Processing,\\
Dhirubhai Ambani Institute of Information and Communication Technology, \\
Near Indroda Circle\\
Gandhinagar, Gujarat, 382007 India\\

\noindent
Telephone: +91 (79) 30510549 \\ 
Fax: : +91 (79) 30520010 \\
E-mail: m.k.gupta@ieee.org

\newpage

\section{Introduction}

There has been a burst of activities and research in codes over finite rings in last decade,
In particular codes over $\Z_{p^{s}}$ and $\Z_{4}$ received much attention \cite{bdho99, bsc95, bsbm97, 
cs93, dhs99, dgh99, gh98, hkcss94, ha98, Z4-shadow, Kerdock, rs98}.
The covering radius of binary linear codes is a widely studied parameter\cite{ckms85,chll97}.  
Recently the covering radius of codes over $\Z_4$ has been investigated with respect to 
Lee and Euclidean distances \cite{aghos99}. 
Several upper and lower bounds on the covering 
radius has been obtained. In this paper we investigate the coverning radius of the 
codes over $\Z_{2^{s}}$. In particular some bounds
of \cite{aghos99} have been generalized for codes over $\Z_{2^s}$. We also investigate the covering
radius of the $\Z_4$ simplex codes (both types) and their duals, MacDonald codes and repetition codes.

A {\em linear code} $\C,$ of length $n$, over $\;\Z_{p^{s}}$ is an additive subgroup of $\;\Z_{p^{s}}^{n}$.
An element of $\C$ is called a {\em codeword of} $\C$ and a {\em generator matrix} of $\C$ is a
matrix whose rows generate $\C$.
The {\em Hamming weight} $w_H(\x)$ of a vector $\x$ in $\ZZ_{p^{s}}^n$ 
is the number of non-zero components.
The {\em Homogeneous weight} $w_{HW}(\x)$ \cite{Chh96} of a vector $\x=(x_1,x_2,\ldots,x_n) \in \Z^n_{2^{s}}$ 
is given by $\sum_{i=1}^n  w_{HW}(x_i)$ where 
\begin{equation}\label{glw}
w_{HW}(x_i)=\left\{\begin{array}{cc}2^{s-2},& x_i \neq 2^{s-1}\\
2^{s-1},& x_i=2^{s-1}.\end{array}\right.
\end{equation}
In particular, for $s=2$, Homogeneous weight $w_{HW}(\x)$ reduces to {\em Lee weight} $w_L(\x)$ given by
$\sum_{i=1}^n \min\{|x_i|,|4-x_i|\}$.
The {\em Euclidean weight} $w_E(\x)$ of a vector $\x \in \Z^n_{2^{s}}$ is
$\sum_{i=1}^n \min\{x_i^2,(2^s-x_i)^2\}$.
The Euclidean weight is useful in connection with lattice
constructions.
The Hamming, Homogeneous / Lee and Euclidean distances $d_H(\x,\y)$, $d_{HW}(\x,\y) / d_L(\x,\y)$ and $d_E(\x,\y)$ 
between two vectors $\x$ and $\y$ are $w_H(\x-\y)$, $w_{HW}(\x-\y) / w_L(\x-\y)$ and $w_E(\x-\y)$, respectively.
The minimum Hamming, Homogeneous / Lee and Euclidean weights, $d_H, d_{HW} / d_L$ and $d_E$, of $\C$ are the smallest Hamming, 
Homogeneoues/ Lee and Euclidean weights among all non-zero codewords of $\C,$ respectively.
One can define an isometry (called {\em Generalized Gray map} \cite{carlet98jul}) from $\left(  \Z_{2^s}, w_{HW} \right)  \rightarrow \left( {\Z^{2^{s-1}}_2}, w_H \right) $
which maps a linear code over $\Z_{2^{s}}$ to a binary code of length $2^{s-1}$ times and with minimum Hamming weight equal to minimum Homogeneous 
weight of pre-image code over $\Z_{2^s}$. In particular, the {\em Gray map} $\phi : \;\Z_4^{n} \rightarrow \;\Z_2^{2n}$
is the coordinate-wise extension of the function from $\;\Z_4$ to $\;\Z_2^{2}$ defined by 
$0 \rightarrow (0,0), 1 \rightarrow (0,1), 2 \rightarrow (1,1), 3 \rightarrow (1,0)$.
The image $\phi(\C)$, of a linear code $\C$ over $\;\Z_4$ of length $n$ by the Gray map, is 
a binary code of length $2n$ \cite{hkcss94}.

The {\em dual code} $\C^{\perp}$ of $\C$ is defined as
$\{ \x \in {\ZZ}_{2^{s}}^n \mid  \x \cdot \y = 0 \;\mbox{for all}\; \y \in \C\}$ 
where $\x \cdot \y$ is the standard inner product of $\x$ and $\y$.
$\C$ is {\em self-orthogonal} if $\C \subseteq \C^\perp$ and $\C$ is
{\em self-dual} if $\C=\C^\perp$.

Two codes are said to be {\em equivalent} if one can be obtained from the
other by permuting the coordinates and (if necessary) changing the signs of
certain coordinates.
Codes differing by only a permutation of coordinates are called
{\em permutation-equivalent}.

In this paper we define the covering radius of codes over $\Z_{2^s}$ with respect to 
different distances and in particular study the covering radius of $\;\Z_4$-simplex codes 
of type $\;\alpha$ and $\beta$ namely, $S_k^{\alpha}$ and $S_k^{\beta}$ and their duals, MacDonald codes and repetition codes.
Section $2$ contains some preliminaries and notations. 
Basic results for the covering radius of codes over $\Z_{2^s}$ are given in Section $3.$
Section $4$ determines the covering radii of different $\Z_4$ repetition codes. Section $5$ determines the covering radius of $\Z_4$ 
Simplex codes and its dual and finally Section $6$ determines the bounds on the covering radius of $\Z_4$ MacDonald codes.

\section{Preliminaries and Notations}

Any linear code $\C$ over $\ZZ_{p^{s}}$ is permutation-equivalent to a code with generator
matrix $G$ (the rows of $G$ generate $\C$) of the form
\begin{equation}\label{eqn:1}
G=\left[\begin{array}{cccccc}I_{k_0}&A_{01}&A_{02}& \cdots&A_{0s-1}&A_{0s}\\
{\bf 0}&pI_{k_1}&pA_{12}& \cdots &pA_{1s-1}& pA_{1s}\\
{\bf 0}&{\bf 0}&p^{2}I_{k_2}& \cdots & p^{2}A_{2s-1} & p^{2}A_{2s}\\
\vdots& \vdots & \vdots & \ddots & \vdots & \vdots \\
{\bf 0}&{\bf 0}&{\bf 0}& \cdots & p^{s-1}I_{k_{s-1}} & p^{s-1}A_{s-1s}
\end{array}\right],
\end{equation}
\noindent where $A_{ij}$ are matrices over $\;\Z_{p^{s}}$ and the columns are
grouped into blocks  of sizes $k_0, \; k_1, \; \cdots, \; k_{s-1}, \; k_{s},$ respectively.
Let $k=\sum_{i=0}^{s-1} (s-i)k_i$. Then $|\C|= p^{k}$.
For $s=2,p=2$, two binary codes (residue and torsion) obtained from code over  $\ZZ_{4}$ are well studied.
For each  $a \in \Z_4 $ let $\bar{a}$ be the reduction of $ a$
modulo $2$ then the code 
$$\C^{(1)}= \left\{(\bar{c_1}, \bar{c_2}, \ldots, \bar{c_n} ) \mid 
(c_1, c_2, \ldots, c_n) \in \C \right\}$$ 
is a binary linear code called the {\em 
residue code} of $\C.$ Another binary linear code associated with $\C$ is the 
{\em torsion code} $\C^{(2)}$ which is defined by 
$$\C^{(2)} = \left\{ \c \in \Z_2^{n} \mid 2 \c \in \C \right\}.$$

  A vector $\v \in \Z^{n}_{p^{s}}$ is a {\em $p$-linear combination} of the vectors $\v_1, \v_2, \ldots, 
\v_k \in \Z^{n}_{p^{s}}$ if $ \v = \l_1 \v_1 + \ldots + \l_k \v_k$ with   $\l_i \in \Z_p$ for $ 1 \leq i 
\leq k.$ A subset $ S = \{ \v_1,\v_2, ...,\v_k \}$ of $\C$ is called a {\em $p$-basis} 
for $\C$ if  for each $i= 1,2,...,k-1, \; p \v_i$ is a $p-$linear combination of $ 
\v_{i+1},..., \v_k$, $ p \v_k = 0, \; \C$ is the $p$-linear span of $S$ and $S$ is 
$p$-linearly independent \cite{vsr96}. The number of elements in a $p$-basis for 
  $\C$ is called the {\em $p$-dimension} of $\C.$
It is easy to verify that the rows of the matrix 
    
 \begin{equation}\label{eqn:2}
\cal{B} = \left[\begin{array}{llllll}
I_{k_0}&A_{01}&A_{02}& \cdots&A_{0s-1}&A_{0s}\\ \hline
pI_{k_0}&pA_{01}&pA_{02}& \cdots&pA_{0s-1}&pA_{0s}\\
{\bf 0}&pI_{k_1}&pA_{12}& \cdots &pA_{1s-1}& pA_{1s}\\ \hline
\vdots & \vdots & \vdots & & \vdots & \vdots \\
\vdots & \vdots & \vdots &  & \vdots & \vdots \\
\hline
p^{s-1}I_{k_0}&p^{s-1}A_{01}&p^{s-1}A_{02}& \cdots&p^{s-1}A_{0s-1}&
p^{s-1}A_{0s}\\
{\bf 0}&p^{s-1}I_{k_1}&p^{s-1}A_{12}& \cdots &p^{s-1}A_{1s-1}&
p^{s-1}A_{1s}\\
{\bf 0}&{\bf 0}&p^{s-1}I_{k_2}& \cdots & p^{s-1} A_{2 s-1} & p^{s-1}A_{2s}\\
\vdots & \vdots & \vdots & \ddots & \vdots & \vdots \\
{\bf 0}&{\bf 0}&{\bf 0}& \cdots & p^{s-1}I_{k_{s-1}} & p^{s-1}A_{s-1s}
\end{array}\right].
\end{equation}

form a $p$-basis for the code $\C$  generated by $G$ given in (\ref{eqn:1}).  
Thus $p\!-\!\dim(\C)= k =\sum_{i=0}^{s-1} (s-i)k_i.$ From now on we restrict to the case 
of $p=2$.
      

A linear code $\C$ over $\;\Z_{2^s}$ ( over $\;\Z_2$) of length $n$,
$2$-dimension $k$, minimum distance $d_H, d_{HW}$ and $d_E$ is called an
$\left[ n,k,d_H,d_{HW},d_E \right]$ $\left([n,k,d_H]\right)$ or simply
an $\left[ n,k \right]$ code. 

\section{Covering Radius of Codes}

In this section, we describe some properties of the covering radius of codes over $\Z_{2^s}$
after giving the definition of the covering radius for the codes over $\Z_{2^s}$. Since for
the codes over $\Z_{2^s}$ various distances are possible we give a definition of the covering 
radius for a general distance which could be any of the possible distance. Let $d$ be the 
general distance out of various possible distances (such as Hamming, Lee, Homogenous and Euclidean). The {\em covering radius} of a code 
$\C$ over $\Z_{2^s}$ with respect to a general distance $d$ is given by
$$r_d(\C)= \max_{\u \in \Z_{2^s}^{n}}\left\{\min_{\c \in \C}d(\u,\c)\right\}.$$    
It is easy to see that $r_d(\C)$ is the minimum value $r_d$ such that 
$$\Z_{2^s}^{n}= \cup_{\c \in \C} S_{r_d}(\c)$$ where
$$S_{r_d}(\u)=\left\{\v \in \Z_{2^s}^{n} \mid d(\u,\v) \leq r_d \right\}$$ for 
any element $\u \in \Z_{2^s}^{n}$.

The translate $\u + \C = \left\{\u + \c \mid \c \in \C \right\}$ is called 
the coset of $\C$ where $\u$ is a vector of $\Z_{2^s}^{n}$. A vector of minimum
weight in a coset is called a {\em coset leader}. The following proposition is
straigthforward generalization from a proposition ~\cite{aghos99}. 

\begin{proposition}
The covering radius of $\C$ with respect to the general distance $d$ is the 
largest minimum weight among all cosets.
\end{proposition}

Also the following proposition is straightforward~\cite{aghos99}.

\begin{proposition}
Let $\C$ be a code over $\Z_{2^s}$ and $\phi(\C)$ the generalized Gray map image of $\C$.
Then $r_{HW}(\C)=r_H(\phi(\C))$.
\end{proposition}

Now we give several lower and upper bounds on the covering radius of codes over $\Z_{2^s}$ with respect 
to homogenous weight. The proof of Proposition $3$ and Theorem $1$, being similar to the case of $\Z_4$ ~\cite{aghos99}, is omitted.

\begin{proposition}{\bf(Sphere-Covering Bound)}
For any code $\C$ of length $n$ over $\Z_{2^s}$, 
$$\frac{2^{2^{s-1} n}}{|\C|} \leq \sum_{i=0}^{r_{HW}(\C)} {2^{s-1} n  \choose i}.$$
\end{proposition}

Now we consider the two upper bounds on the covering radius of a code over $\Z_{2^s}$
with respect to homogenous weight. Let $\C$ be a code over $\Z_{2^s}$ and let 
$$s(\C^{\perp})=|\left\{i \mid A_i(\C^{\perp}) \neq 0, i \neq 0 \right\}|$$ 
where $A_i(\C^{\perp})$ is the number of codewords of homogenous weight $i$ 
in $\C^{\perp}$. 

\begin{theorem}{\bf (Delsarte Bound)}
Let $\C$ be a code over $\Z_{2^s}$ then $r_{HW}(\C) \leq s(\C^{\perp})$.
\end{theorem}


The following result of Mattson \cite{ckms85} is useful  for computing covering radii of codes over rings generalized easily
from codes over finite fields.  

\begin{proposition} {\bf (Mattson)}\label{mattson}
If $\C_0$ and $\C_1$ are codes over $\Z_{2^s}$ generated by matrices $G_0$ and $G_1$ respectively and 
if $\C$ is the code generated by 
\[
G =  \left( \begin{array}{c|c} 
0 & G_1 \\\hline
G_0 & A 
\end{array}\right),
\]
then $r_d(\C) \leq r_d(\C_0) + r_d(\C_1)$ and  the covering radius of $\D$ (concatenation of $\C_0$ and $\C_1$) 
satisfy the following 
\[
r_d(\D) \geq r_d(\C_0) + r_d(\C_1), 
\]
for all distances $d$ over $\Z_{2^s}$.

\end{proposition}

\section{Repetition Codes}
A $q$-ary repetition code $\C$ over a finite field $\FF_q= \{\alpha_0 = 0,\alpha_1 =1, \alpha_2, \alpha_3, \ldots, \alpha_{q-2} \}$ is
an $[n,1,n]$ code $\C = \{ \bar{\alpha} | \alpha \in \FF_q \},\;\mbox{where}\; \bar{\alpha} = (\alpha, \alpha, \ldots, \alpha)$. The covering 
radius of $\C$ is $\lceil \frac{n(q-1)}{q} \rceil$ \cite{duraithesis96}. Using this it can be seen easily 
that the covering radius of block (of size $n$) repetition code $[n(q-1),1,n(q-1)]$ generated by 
$G= [ \overbrace{ 11 \ldots 1}^{n} \overbrace{\alpha_2 \alpha_2 \ldots \alpha_2}^{n} \ldots \overbrace{\alpha_{q-2} \alpha_{q-2} \ldots \alpha_{q-2}}^{n} ]$ is  $\lceil \frac{n(q-1)^2}{q} \rceil$
(since it will be equivalent to a repetition code of length $(q-1)n$).  

Consider the repetition code over $\Z_4$. There are two
types of them of length $n$ viz. unit repetition code $\C_{\beta}: [n,2,n,n]$ generated by $G_{\beta}=[\overbrace{1 1 \ldots 1}^{n}]$ and zero divisor repetition code $\C_{\alpha}: [n,1,n,2n]$ generated by $G_{\alpha}=[\overbrace{2 2 \ldots 2}^{n}]$. The following result determines the covering radius for both.

\begin{theorem}
$r_L(\C_{\alpha})=n, r_E(\C_{\alpha})=2n, r_L(\C_{\beta})=n\; \mbox{and}\; r_E(\C_{\beta})=\frac{3n}{2}.$
\end{theorem}

\begin{proof}
Note that $\phi(\C_{\alpha})$ is a binary repetition code of length $2n$ hence $r_L(\C_{\alpha})=\frac{2n}{2}=n$. 
Now  by definition $r_E(\C_{\alpha}) = \max_{\x \in \Z^n_4} \{ d_E(\x,\C_{\alpha}) \}$. Let $\x =\overbrace{2 2 2 \ldots 2}^{\frac{n}{2}}\overbrace{0 0 0 \ldots 0}^{\frac{n}{2}} \in \Z^n_4$, then 
$d_E(\x,\bar{0})=d_E(\x,\bar{2}) = 2n$. Thus $ r_E(\C_{\alpha}) \geq 2n$. On the other hand if $\x \in \Z^n_4$ has a composition $(\omega_0, \omega_1, \omega_2, \omega_3)$, where 
$\sum_{i=0}^{3} \omega_i = n$ then $d_E(\x,\bar{0})=n-\omega_0+3 \omega_2$ and  $d_E(\x,\bar{2})=n-\omega_2+3 \omega_0$. Thus $d_E(\x,\C_{\alpha}) = \min \{ n-\omega_0+3 \omega_2, n-\omega_2+3 \omega_0 \} \leq n+\omega_0 + \omega_2 \leq n+n = 2n$. Hence  $r_E(\C_{\alpha})=2n.$ Similar arguments can be used to show that $r_E(\C_{\beta}) \leq  \frac{3n}{2}.$ To show that 
$r_E(\C_{\beta}) \geq  \frac{3n}{2}, $ let $\x=\overbrace{0 0 0 \ldots 0}^{t}\overbrace{1 1 1 \ldots 1}^{t}\overbrace{2 2 2 \ldots 2}^{t}\overbrace{3 3 3 \ldots 3}^{n-3t} \in  \Z^n_4$, where $t=\lfloor \frac{n}{4} \rfloor$, 
then $d_E(\x,\bar{0})=n+2t, d_E(\x,\bar{1})=4n-10t, d_E(\x,\bar{2})=n+2t$ and $d_E(\x,\bar{3}) =6t$. Thus $ r_E(\C_{\beta}) \geq \min \{4n-10t,n+2t,6t \} \geq \frac{3n}{2}$.  Thus $r_E(\C_{\beta})=\frac{3n}{2}$.
The proof of $r_L(\C_{\beta})=n$ is simple so we omit it.
\end{proof}   
\qed

In order to determine the covering radius of Simplex and MacDonald codes over $\Z_4$, we need to define few block repetition codes over $\Z_4$ and find their covering radii.  
To determine the covering radius of  $\Z_4$  block (three blocks each of size $n$) repetition code $BRep^{3n}_{\alpha}: [3n,2,2n,4n,6n]$ generated by 
$G= [ \overbrace{11 \ldots 1}^{n}\overbrace{2 2 \ldots 2}^{n} \ldots \overbrace{3 3 \ldots 3}^{n} ]$ note 
that the code has constant Lee weight $4n$. Thus  for  $\x= 11 \ldots 1 \in {\Z^{3n}_4}$, we have
$d_{L}(\x,BRep^{3n}_{\alpha}) = 3n$. Hence by definition, $r_L({BRep^{3n}_{\alpha}}) \geq 3n$. On the other hand, its Gray image $\phi(BRep^{3n}_{\alpha})$ is equivalent to binary linear code $[6n,2,4n]$ with the generator matrix 
\[
\left(\begin{array}{c|c|c}
\overbrace{1 1 \ldots 1}^{2n} &\overbrace{1 1 \ldots 1}^{2n}&  \overbrace{0 0 \ldots 0}^{2n} \\
\underbrace{1 1 \ldots 1}_{2n}& 0 0 \ldots 0 & 1 1 \ldots 1 \\
\end{array} \right) .
\]
Thus the covering radius $r_L({BRep^{3n}_{\alpha}}) \leq \frac{4n}{2} + \frac{2n}{2} = 3n$. This completes the proof of the first part of  useful Theorem ~\ref{brep3n}. For the second part note that 
$r_E({BRep^{3n}_{\alpha}})  \geq \frac{3n}{2} + 2n + \frac{3n}{2}  = 5n.$ To find an upper bound let $\x= (\u | \v |  \w) \in {\Z^{3n}_4}$, with $\u, \v$ and $\w$ have compositions $(r_0,r_1,r_2,r_3)$, $(s_0,s_1,s_2,s_3)$ and 
 $(t_0,t_1,t_2,t_3)$ respectively such that sum of each component composition is $n$, then $d_E(\x,\bar{0})= 3n-r_0+3r_3-s_0-3s_3-t_0+3t_3,  d_E(\x,\c_1) = 3n -r_1+3r_0-s_2+3s_1-t_3+3t_2, 
  d_E(\x,\c_2) = 3n-r_2+3r_1-s_0+3s_3-t_2+3t_1$ and   $d_E(\x,\c_3)= 3n-r_3+3r_2-s_2+3s_1-t_1+3t_0$. Thus      $d_E(\x,{BRep^{3n}_{\alpha}})    \leq 3n + \min \{ 3r_3+3s_3+3t_3-r_0-s_0-t_0,  3r_0+3s_2+3t_2-r_1-s_2-t_3, 3r_1+3s_3+3t_1-r_2-s_0-t_2, 3r_2+3s_1+3t_0-r_3-s_2-t_1 \} \leq 3n + \frac{1}{2} \{ n+4s_1+4s_3 \} \leq \frac{11n}{2} $. 

\begin{theorem}\label{brep3n}\label{brep3n}
$r_L({BRep^{3n}_{\alpha}}) = 3n$ and $5n \leq r_E({BRep^{3n}_{\alpha}}) \leq \frac{11n}{2}.$
\end{theorem}

One can also define a  $\Z_4$  block (two blocks each of size $n$) repetition code $BRep^{2n}_{\alpha}: [2n,2,n,2n,4n]$ generated by 
$G= [ \overbrace{11 \ldots 1}^{n}\overbrace{2 2 \ldots 2}^{n}]$. We have following theorem (its proof is similar to the proof of Theorem \ref{brep3n}) so we omit it.

\begin{theorem}\label{brep2n}
$r_L({BRep^{2n}_{\alpha}}) = 2n$ and $r_E({BRep^{2n}_{\alpha}}) = \frac{7n}{2}.$
\end{theorem}

Block code $BRep^{2n}_{\alpha}$ can be generalized to a block repetition  code (two blocks of size $m$ and $n$ respectively) $BRep^{m+n}: [m+n,2,m, \min \{2m,m+2n\}, \min \{4m, m+4n\}]$
generated by $G= [ \overbrace{11 \ldots 1}^{m}\overbrace{2 2 \ldots 2}^{n}]$. Theorem \ref{brep2n} can be easily generalized using similar arguments to the following.

\begin{theorem}\label{brep_mn}
$r_L({BRep^{m+n}}) = m+n$ and $ r_E({BRep^{m+n}}) = 2n+\frac{3m}{2} .$
\end{theorem}


\section{Quaternary Simplex Codes of Type $\alpha$ and $\beta$}

Quaternary simplex codes of type $\alpha$ and $\beta$ have been recently studied 
in \cite{bgl99}. Type $\alpha$ simplex code $S_k^{\alpha}$ is a linear code over
$\Z_4$ with parameters $\left[2^{2k},2k,2^{2k-1},2^{2k},3 \cdot 2^{2k-1}\right]$ 
and an inductive generator matrix given by 
\be \label{skalpha}
G_k^{\alpha} = \left[\begin{array}{c|c|c|c} 0\; 0 \cdots 0 & 1\; 1 \cdots 1 &
2\; 2 \cdots 2 & 3\; 3 \cdots 3 \\\hline
G_{k-1}^{\alpha} & G_{k-1}^{\alpha} & G_{k-1}^{\alpha}&G_{k-1}^{\alpha}
\end{array}\right]\ee
with $G_1^{\alpha}$ =$[ 0\; 1\; 2\; 3 ]$.
The dual code of $S_k^{\alpha}$ is a $\left[2^{2k},2^{2k+1}-2k\right]$ code.
Type $\beta$ simplex code $S_k^{\beta}$ is a punctured version of $S_k^{\alpha}$
with parameters $$\left[2^{k-1}(2^k-1),2k,2^{2(k-1)},2^{k-1}(2^k-1),2^k(3 \cdot 2^{k-2}-1)\right]$$
and an inductive generator matrix given by
\be \label{skbeta2}
G_2^{\beta} = \left[\begin{array}{c|c|c} 1\; 1\; 1\; 1 & 0 & 2 \\\hline
0\; 1\; 2\; 3 & 1 & 1\end{array}\right], \ee
and for $k > 2$
\be \label{skbetak}
G_k^{\beta} = \left[\begin{array}{c|c|c} 1\; 1 \cdots 1 & 0\; 0 \cdots 0
& 2\; 2 \cdots 2 \\\hline
G_{k-1}^{\alpha} & G_{k-1}^{\beta} & G_{k-1}^{\beta}\end{array}\right], \ee 
where $G_{k-1}^{\alpha}$ is the generator matrix of $S_{k-1}^{\alpha}$. For details the 
reader is refereed to \cite{bgl99}. 
The dual code of $S_k^{\beta}$ is a $\left[2^{k-1}(2^k-1),2^{2k}-2^{k}-2k\right]$ type
$\alpha$ code with minimum Lee weight $d_L=3$.  

\begin{theorem}
$r_L({S_k^{\alpha}})=2^{2k}\;\mbox{and}\;r_E({S_k^{\alpha}}) \leq  \frac{11(4^k-1)+9}{6}.$
\end{theorem}

\begin{proof}
Let $\x= 11 \ldots 1 \in \Z^{n}_4$. Since $S_k^{\alpha}$ is of constant Lee weight $(=2^{2k})$ code, we have
$d_{L}(\x,S_k^{\alpha}) = 2^{2k}$. Hence by definition, $r_L({S_k^{\alpha}}) \geq 2^{2k}$.  On the other hand by equation  (\ref{skalpha}), 
the result of Mattson (see Proposition \ref{mattson}) for finite rings and using Theorem \ref{brep3n}, we get
\[
\begin{array}{ccc}
 r_L({S_k^{\alpha}}) & \leq  & r_L({S_{k-1}^{\alpha}}) + r_L( <  \overbrace{1 1 \ldots 1}^{2^{2(k-1)}}\overbrace{2 2 \ldots 2}^{2^{2(k-1)}}  \overbrace{ 3 3 \ldots 3}^{2^{2(k-1)}} >)\\
  & = & r_L({S_{k-1}^{\alpha}}) + 3.2^{2(k-1)} \\
  & \leq & 3.2^{2(k-1)} + 3.2^{2(k-2)} + 3.2^{2(k-3)} + \ldots + 3.2^{2.1} +  r_L({S_{1}^{\alpha}}) \\
  & \leq & 3 (4^{k-1}+4^{k-2}+ \ldots+4+1) +1 (\mbox{since}\; r_L({S_{1}^{\alpha}})=4)\\
  &=& 2^{2k}.
\end{array}
\]
Thus $r_L({S_k^{\alpha}})=2^{2k}$. Similar arguments can be used to show that (using Theorem \ref{brep3n})
\[
\begin{array}{ccc}
r_E({S_k^{\alpha}})   &   \leq & \frac{11}{2}\left( 4^{(k-1)} + 4^{(k-2)} + 4^{(k-3)} + \ldots + 4^{1} +1\right)-\frac{11}{2}+  r_E({S_{1}^{\alpha}}) \\
  & \leq & \frac{11}{6} (4^k-1) -\frac{11}{2}+7\; (\;\mbox{since}\; r_E({S_{1}^{\alpha}})  \leq 7)\\
  && = \frac{11(4^k-1)+9}{6}.
\end{array}
\]

\end{proof}   
\qed

Similar arguments will compute the covering radius of Simplex codes of type $\beta$. We provide an outline of the proof.

\begin{theorem}
$r_L({S_k^{\beta}}) \leq 2^{k-1}(2^k-1)-2\;\mbox{and}\;r_E({S_k^{\beta}}) \leq 2^k(2^{k+1}-1)+\frac{1}{3}(4^k-1)-\frac{147}{2}.$
\end{theorem}

\begin{proof}
By equation  (\ref{skbetak}), Proposition \ref{mattson} and Theorem \ref{brep_mn}, we get
\[
\begin{array}{ccc}
 r_L({S_k^{\beta}}) & \leq  & r_L({S_{k-1}^{\beta}}) + r_L( <  \overbrace{1 1 1 \ldots 1}^{4^{(k-1)}} \overbrace{2 2 2 \ldots 2}^{2^{(2k-3)}-2^{(k-2)}} >)\\
  & = & r_L({S_{k-1}^{\alpha}}) + 2^{(2k-2)}+2^{(2k-3)}-2^{(k-2)} \\
  & \leq & (2^{(2k-2)}+2^{(2k-3)} +\ldots +2^6+2^5+2^4+2^3) - (2^{(k-3)}+2^{(k-4)} + \ldots +2^2+2)+ r_L({S_{2}^{\beta}}) \\
  & \leq &  (2^{(2k-1)}-1)-(2^2+2+1)-(2^{(k-1)}-1)-1+ 6 (\mbox{since}\; r_L({S_{2}^{\beta}}) \leq 7)\\
  &=& 2^{k-1}(2^k-1)-2.
\end{array}
\]
Thus $r_L({S_k^{\beta}}) \leq 2^{k-1}(2^k-1)-2$. Similar arguments can be used to show that (using Theorem \ref{brep3n})
\[
\begin{array}{ccc}
r_E({S_k^{\beta}})   &   \leq &  2^{(k-1)}(2^{(k-1)}-1)+2^{(k-2)}(2^{(k-2)}-1)+ \ldots +2^3(2^3-1)+2^2(2^2-1)\\
& &+3 (2^{(2k-1)}+2^{(2k-3)}+\ldots + 2^7+2^5)+  r_E({S_{2}^{\beta}}) \\
  & \leq&2^{2k+1}+ \frac{1}{3} (4^k-1)-(2^k-1)-4^3-4^2-4 + \frac{19}{2} \; (\;\mbox{since}\; r_E({S_{2}^{\beta}})  \leq \frac{19}{2})\\
  && = 2^k(2^{k+1}-1)+\frac{1}{3}(4^k-1)-\frac{147}{2}.
\end{array}
\]

\end{proof}   
\qed

\begin{theorem}
$r_L({S_k^{\alpha}}^{\perp})=1$, $r_L({S_k^{\beta}}^{\perp}) = 2,$ $r_E({S_k^{\alpha}}^{\perp}) \leq 4$ and $ r_E({S_k^{\beta}}^{\perp}) \leq 4$.
\end{theorem}

\begin{proof}
By Delsarte bound, $r_L({S_k^{\alpha}}^{\perp}) \leq 1$ and $r_L({S_k^{\beta}}^{\perp}) \leq 2$.
Thus equality follows in the first case. For second case, note that $r_L({S_k^{\beta}}^{\perp}) \neq 1,$  by sphere-covering bound. The results for 
Euclidean distance follows from  Delsarte bound.
\end{proof}   
\qed

\section{Quaternary MacDonald Codes of Type $\alpha$ and $\beta$}

The $q$-ary MacDonald code $\M_{k,u}(q)$
over the finite field $\FF_q$ is a unique $\left[\frac{q^{k}-q^{u}}{q-1},k,
q^{k-1}-q^{u-1}\right]$ code in which every nonzero codeword has
weight either $q^{k-1}$ or $q^{k-1}-q^{u-1}$ \cite{dodsim98}.
In \cite{cg03}, authors have defined  the MacDonald codes over $\Z_4$ using the generator matrices of
simplex codes.  For $1 \leq u \leq k-1,$ let
$G_{k,u}^{\alpha}\left(G_{k,u}^{\beta}\right)$ be the matrix
obtained from $G_{k}^{\alpha}\left(G_{k}^{\beta}\right)$ by
deleting columns corresponding to the columns of
$G_{u}^{\alpha}\left(G_{u}^{\beta}\right)$. i.e, \be
\label{macalpha}
G_{k,u}^{\alpha}=\left[\begin{array}{cc}G_k^{\alpha}& \backslash\;
\frac{\bf{0}}{G_u^{\alpha}}
\end{array} \right],
\ee
and\\
\be \label{macbeta}
G_{k,u}^{\beta}=\left[\begin{array}{cc}G_k^{\beta}&
\backslash\; \frac{\bf{0}}{G_u^{\beta}}
\end{array} \right],
\ee
where $[A \backslash B]$ denotes the matrix obtained
from the matrix $A$ by deleting
the columns of the matrix $B$ and ${\bf 0}$ in
$(\ref{macalpha})\left(\;\mbox{resp.}(\ref{macbeta})\right)$
is a $(k-u) \times 2^{2u}\left(\;\mbox{resp.}\;(k-u)
\times 2^{u-1}(2^{u}-1)\right)$ zero matrix.

The code $\M_{k,u}^{\alpha}:[2^{2k}-2^{2u},2k] \left(\M_{k,u}^{\beta}: [(2^{k-1}-2^{u-1})(2^k+2^u-1),2k]\right)$
generated by the matrix
$G_{k,u}^{\alpha}\left(G_{k,u}^{\beta}\right)$ is the punctured
code of $S_k^{\alpha}\left(S_k^{\beta}\right)$ and is  called a
{\em MacDonald code} of type $\alpha\; ( \beta)$.

Next theorems provides basic bounds on the covering radii of MacDonald codes.

\begin{theorem}
\[
\begin{array}{ccc}
r_L(\M_{k,u}^{\alpha}) & \leq & 4^k-4^r + r_L(\M_{r,u}^{\alpha})\;\mbox{for}\; u < r \leq k,\\
r_E(\M_{k,u}^{\alpha}) &\leq & \frac{11}{6} (4^k-4^{r})+ r_E(\M_{r,u}^{\alpha})\;\mbox{for}\; u < r \leq k.
\end{array}
\]
\end{theorem}

\begin{proof}
By Theorem \ref{brep3n}, 
\[
\begin{array}{ccc}
r_L(\M_{k,u}^{\alpha}) &\leq & 3.2^{(2k-2)} + r_L(\M_{k-1,u}^{\alpha})\\
& \leq & 3.2^{(2k-2)} + 3. 2^{(2k-4)} + \ldots + 3.2^r + r_L(\M_{r,u}^{\alpha}), k \geq r > u\\
 & = &4^k-4^r + r_L(\M_{r,u}^{\alpha}).
\end{array}
\]
Similar arguments holds for $r_E(\M_{k,u}^{\alpha})$.
\end{proof}
\qed

Similarily using equation (\ref{macbeta}), Proposition \ref{mattson} and Theorem \ref{brep_mn} following  bounds can be obtained  for type $\beta$ MacDonald code.

\begin{theorem}
\[
\begin{array}{ccc}
r_L(\M_{k,u}^{\beta}) & \leq & 2^{k-1}(2^k-1) - 2^{r-1}(2^r-1) + r_L(\M_{r,u}^{\beta})\;\mbox{for}\; u < r \leq k,\\
r_E(\M_{k,u}^{\beta}) &\leq & \frac{2^{2r-1}}{3}(4^{k-r+1}-1)+4^{r-1}(4^{k-r}-1)- 3. 2^{r-2}(2^{k-r}-1)+ r_E(\M_{r,u}^{\beta})\;\mbox{for}\; u < r \leq k.
\end{array}
\]
\end{theorem}

\section{Conclusion}
We have computed bounds on the covering radii of Simplex and MacDonald codes over $\Z_4$ and also provided exact values in some cases. It would be 
an interesting future task to find out the exact covering radii of many of these codes and generalize the results for codes over $\Z_{2^s}.$

\bigskip
\noindent
{\bf Acknowledgement.}
The authors would like to thank Patrick Sol\'e for reading the first draft of the paper and pointing out an error in it.
 

\end{document}